\documentclass[useAMS,usenatbib,usegraphicx]{mn2e}
\usepackage{graphicx}                                            
\usepackage{times}
\usepackage{float}
\usepackage{rotating}
\usepackage{epstopdf}
\usepackage{multirow}
\usepackage{pdflscape}
\usepackage{color}

\def\ltsima{$\; \buildrel < \over \sim \;$}
\def\simlt{\lower.5ex\hbox{\ltsima}}
\def\gtsima{$\; \buildrel > \over \sim \;$}
\def\simgt{\lower.5ex\hbox{\gtsima}}
\def\gsimeq
{\hbox{\raise0.5ex\hbox{$>\lower1.06ex\hbox{$\kern-1.07em{\sim}$}$}}}
\def\lsimeq
{\hbox{\raise0.5ex\hbox{$<\lower1.06ex\hbox{$\kern-1.07em{\sim}$}$}}}

\def\chandra{{\it Chandra}}

\def\nustar{{\it NuSTAR}}

\def\athena{{\it Athena}}
\def\xarm{{\it XARM}}
\def\arcus{{\it Arcus}}

\def\apj{ApJ}
\def\mnras{MNRAS}
\def\aap{A\&A}
\def\apjl{ApJ}
\def\apjs{ApJS}
\def\araa{ARA\&A}

\def\iaucirc{IAU~circular}
\def\procspie{Proc. SPIE}

\def\ssr{SSRv}

\def\pasp{PASP}

\def\mxb{MXB~1659-298}

\def\Fevc{Fe~{\sc xxv}}

\def\xis{XIS}
\def\xis1{XIS1}
\def\xis2{XIS2}
\def\xis3{XIS3}

\title[] 
 {{Measuring masses in low mass X-ray binaries via X-ray spectroscopy: 
 the case of \mxb}} 
 \author[G.\ Ponti et al. ]
 {Gabriele~Ponti$^{1}$, Stefano Bianchi$^{2}$, Teo Mu\~{n}oz-Darias$^{3,4}$ 
 and Kirpal Nandra$^{1}$ \\
   $^1$ Max-Planck-Institut f{\"u}r Extraterrestrische Physik, Giessenbachstrasse, D-85748, Garching, Germany\\
   $^2$ Dipartimento di Matematica e Fisica, Universit\`a degli Studi Roma Tre, Via della Vasca Navale 84, I-00146, Roma, Italy\\
   $^{3}$ Instituto de Astrof\'isica de Canarias, 38205 La Laguna, Tenerife, Spain \\
$^{4}$ Departamento de astrof\'isica, Univ. de La Laguna, E-38206 La Laguna, Tenerife, Spain \\
      }
\pagerange{\pageref{firstpage}--\pageref{lastpage}}
\usepackage{times}
\begin{document}
\label{firstpage}
 \maketitle
\begin{abstract}
The determination of fundamental parameters in low-mass X-ray binaries 
typically relies on measuring the radial velocity 
curve of the companion star through optical or near-infrared spectroscopy. 
It was recently suggested that high resolution X-ray spectroscopy 
might enable a measurement of the radial velocity curve of the compact object 
by monitoring the Doppler shifts induced by the orbital motion of the disc wind 
or the disc atmosphere. We analysed a \chandra-HETG+\nustar\ soft state observation 
of \mxb, an eclipsing neutron star low-mass X-ray binary (LMXB). 
We measured a radial velocity curve whose phase offset and semi-amplitude 
are consistent with the primary star. 
We derived the value for the semi-amplitude of the radial velocity for the compact 
object $K_1=89\pm19$~km~s$^{-1}$, constrained the mass of 
the secondary ($0.3$$\leq$$M_2$$\leq$$0.8$~M$_\odot$) and the orbital inclination 
of the binary system ($73$$\leq$$i$$\leq$$77^\circ$). 
These values are consistent with previous estimates from independent methods.
Via the same technique, the next generation of X-ray observatories equipped 
with high spectral resolution instruments (e.g., \athena) will have the potential 
to measure the radial velocity curve of the primary in high inclination X-ray binaries 
to an accuracy of a few per cent. 
\end{abstract}


\section{Introduction}

Dynamical masses of black holes and neutron stars (NS) in X-ray 
binaries (XRB) can be derived through phase-resolved photometric 
and spectroscopic campaigns in the optical or the near infrared 
(nIR; Cowley et al. 1992; Charles \& Coe 2006, Casares \& Jonker 2014). 
This is achieved by measuring the Doppler 
motion and rotational broadening of the absorption lines generated 
at the photosphere of the companion star. These quantities provide, 
under the assumption of co-rotation, a determination of the mass 
function, the mass ratio and the binary orbital inclination 
(e.g. Casares \& Jonker 2014). Decades of 
optical-nIR spectroscopic studies have demonstrated the power 
of these techniques to determine the masses of compact objects, 
however they are limited to sources with relatively bright 
companion stars in quiescence or the presence of fluorescence lines 
from the donor in outburst (Steeghs \& Casares 2002; 
Mu\~noz-Darias et al. 2005). However, often the semi-amplitude 
of the radial velocity of the compact object ($K_\mathrm{1}$) remains 
elusive. Though, $K_\mathrm{1}$ is fundamental to obtain a dynamical 
solution, whenever the mass ratio of the two stars cannot be determined, 
and it is of great value for searches of continuous gravitational waves 
(Watts et al. 2008). 

It was recently proposed an alternative technique to determine 
the mass of compact objects, based on high spectral resolution 
X-ray spectroscopy (Zhang et al. 2012). This method relies on measuring 
orbital shifts in the observed energies of absorption lines from 
disc winds or disc atmospheres, since these are expected 
to trace the motion of the compact object around the centre of mass 
of the binary system (i.e. $K_\mathrm{1}$).
This technique has been already applied 
to a few black hole and NS systems (Zhang et al. 2012; Madej et al. 2014). 
However, it was concluded that the variability of either 
the source luminosity or random variations of the wind 
outflow speed can severely affect the accurate determination of the orbital motion 
of the compact object (Madej et al. 2014). 
In addition, for typical low-mass X-ray binaries (LMXB), 
the expected radial velocity of the primary is of the order 
of $K_1\sim10-150$~km~s$^{-1}$, 
therefore beyond the energy resolution of current X-ray 
instruments in the Fe~K band, where the strongest absorption 
lines are present (Ponti et al.\ 2012; Diaz-Trigo et al.\ 2013). 

Here, we report on the determination of the radial velocity curve 
of the primary in \mxb. This was achieved by applying the method 
proposed by Zhang et al. (2012) to several absorption lines 
in the soft X-ray band, where current X-ray instruments provide 
the highest energy resolution. \mxb\ is a transient LMXB displaying 
type-I X-ray bursts, therefore indicating a neutron star primary 
(Lewin et al. 1976; Galloway et al. 2008). It is a high inclination 
system, showing dipping and eclipsing events, with an orbital 
period of $P_{orb}=7.1$~hr and an eclipse duration of $\approx$$900$~s 
(Cominsky 1984; Jain et al 2017; Iaria al 2018). 

The optical counterpart of \mxb\ was found to have
$V\sim18$ during outburst (Doxsey et al. 1979) and 
to display orbital brightness variations as well as narrow 
eclipses (Wachter et al. 2000). 
The optical spectrum during outbursts is rather typical 
for a LMXB, with a blue continuum, He~{\sc ii} and Bowen blend 
emission but no spectral features from the donor (Canizares et al. 1980; 
Shahbaz et al. 1996). During quiescence, \mxb\ is very faint 
with $V>23$ (Cominsky et al. 1983), $R=23.6\pm0.4$ and 
$I=22.1\pm0.3$ (Filippenko et al. 1999; Wachter et al. 2000). 
Assuming a reddening of $E_{B-V}=0.3$ (van Paradijs \& McClintock 1995) 
and based on the observed 
$(R-I)_0=1.2$ and the empirical period-mass relation, 
Wachter et al. (2000) suggested that the companion star is an early K 
to early M main sequence star. 

\mxb\ started a new outburst on August 2015 (Negoro et al. 2015). 
LMXB observed at high inclination, such as \mxb, are known 
to display strong ionised absorption during the soft state, associated 
with an equatorial wind or the disc atmosphere (Diaz-Trigo et al. 2006; 
Ponti et al. 2012). 
Therefore, in order to perform the first high energy 
resolution study of the ionised absorber in \mxb, 
we triggered \chandra\ and simultaneous \nustar\ observations. 
\begin{figure}
\hspace{-0.3cm}
\includegraphics[height=0.39\textwidth,angle=0]{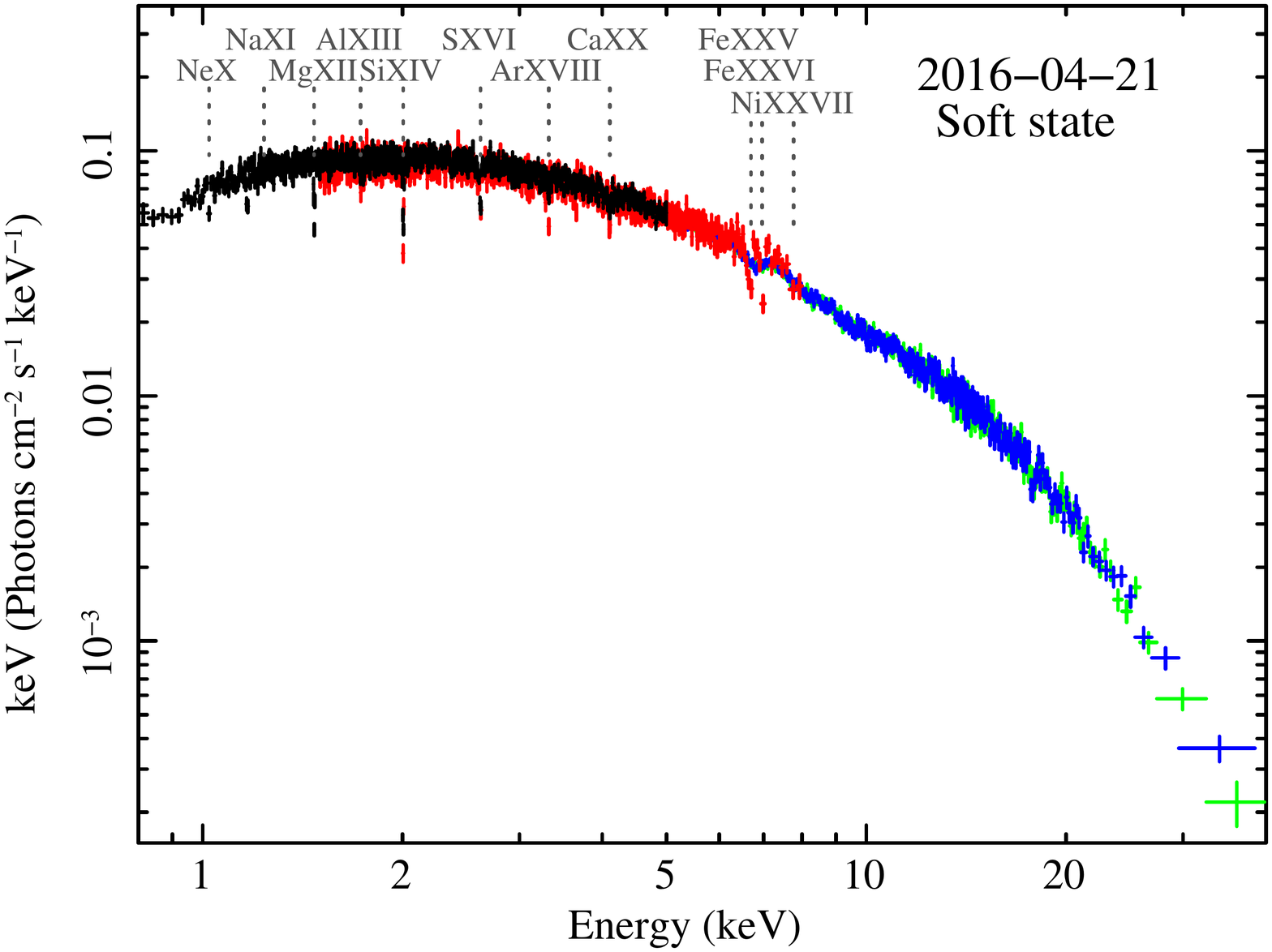}

\vspace{-4.6cm}
\hspace{1.12cm}
\includegraphics[height=0.195\textwidth,angle=0]{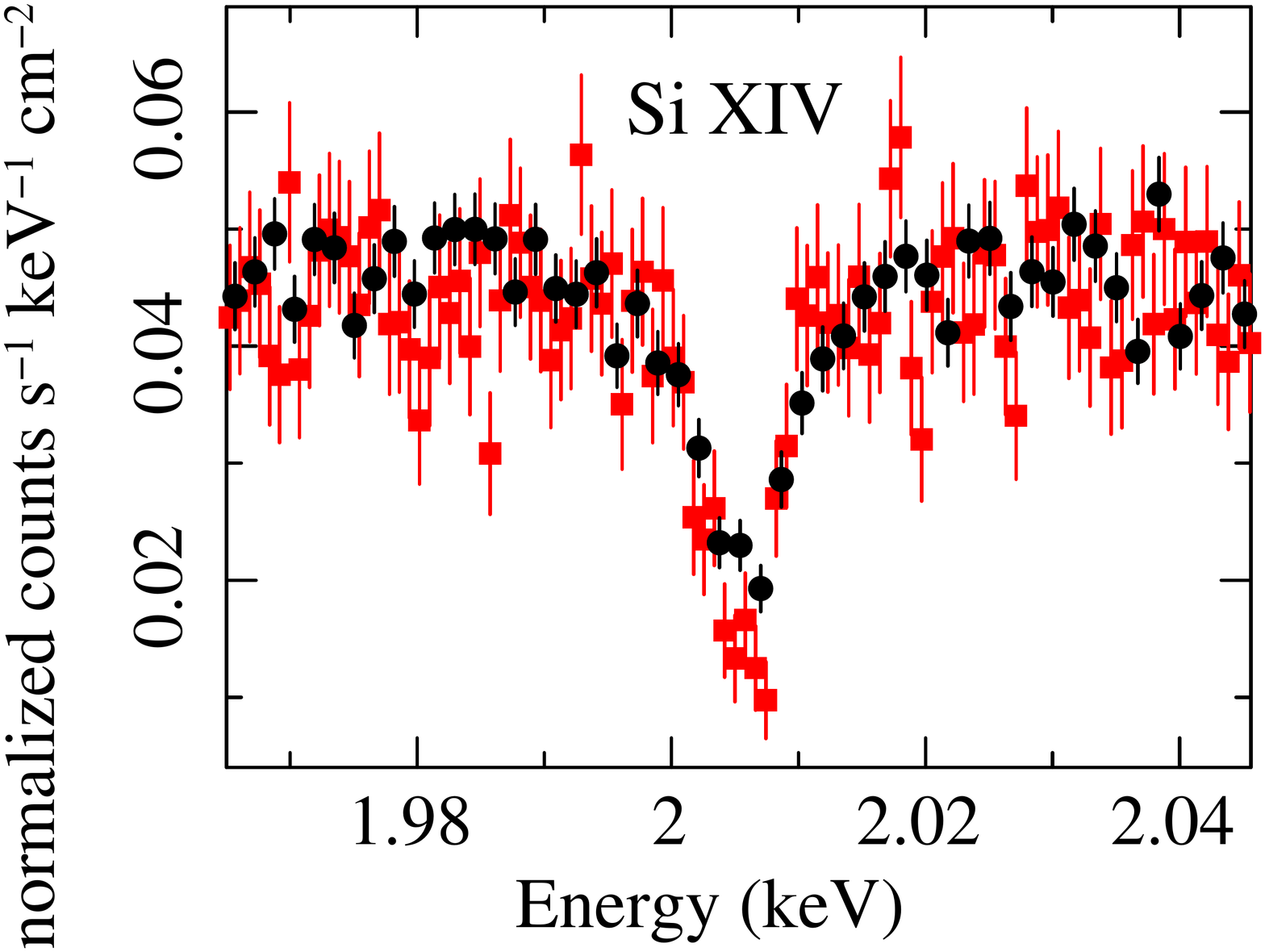}

\vspace{0.8cm}
\caption{\chandra+\nustar\ spectra of \mxb, accumulated 
on 2016-04-21. The black, red, green and blue points show the MEG, 
HEG, FPMA and FPMB mean spectra, respectively. As is typical 
for the soft state, the spectra are best fitted with an absorbed disk black-body, 
a black-body and a Comptonisation component (Ponti et al. in prep). 
Additionally, more than 60 absorption lines are detected, signatures 
of an additional ionised absorption component (the strongest Ly$\alpha$ 
lines detected are indicated, as well as the \Fevc\ line). The inset shows 
a zoom into the Si~{\sc xiv} Ly$\alpha$ line. }
\label{spec}
\end{figure}

\vspace{-0.5cm}
\section{Analysis} 

\begin{figure}
\vspace{-0.7cm}
\hspace{-0.5cm}
\includegraphics[height=0.4\textwidth,angle=0]{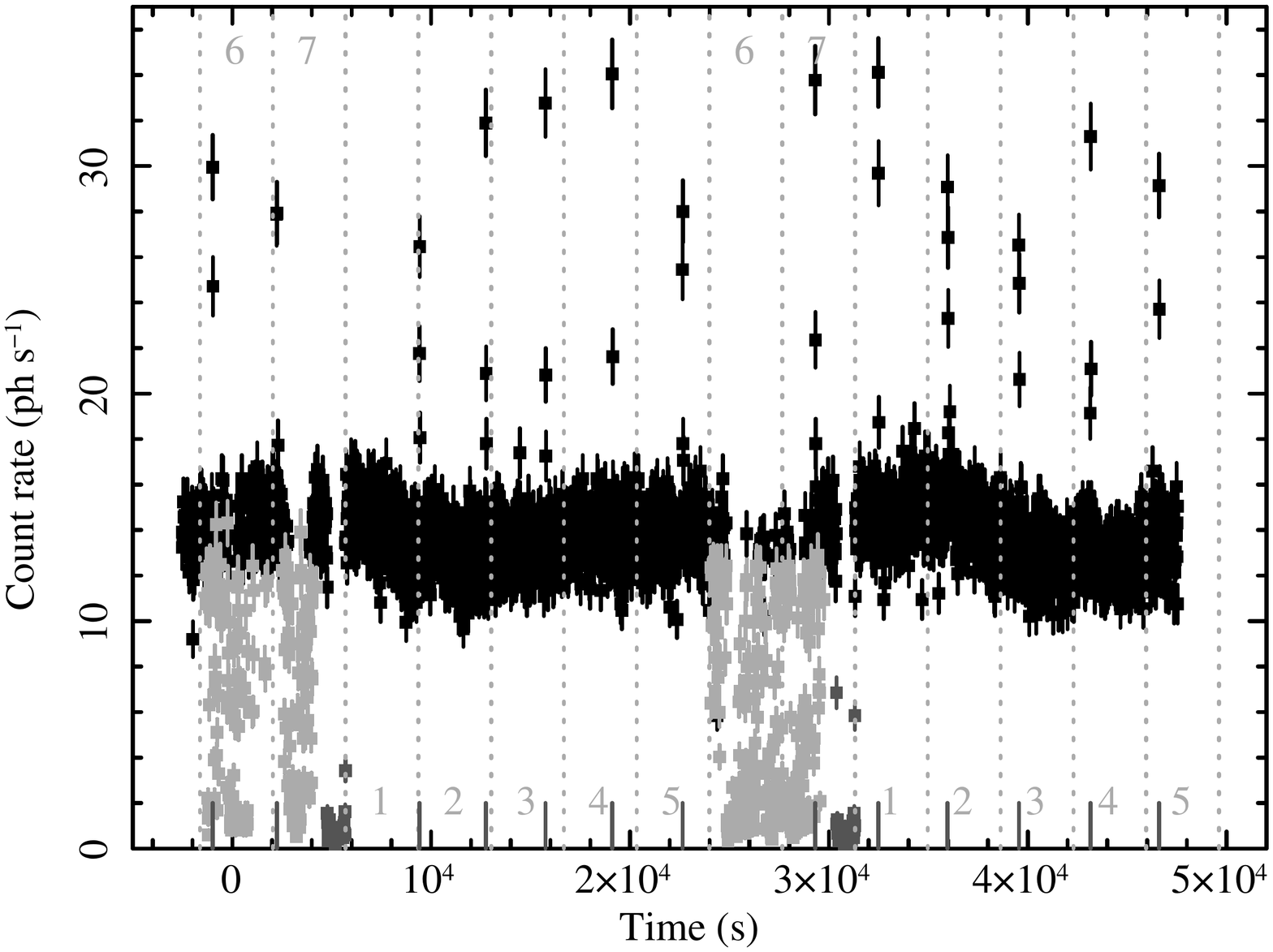}

\vspace{-0.3cm}
\caption{First order \chandra\ HETG light curves of \mxb\ 
($14.9$~s time bins). Intense dipping activity (light grey) is observed, 
primarily before the two eclipses (dark grey). 
Thirteen bursts are also observed (affected by pile up at the burst peaks).
The grey dotted vertical lines indicate the intervals used for the phase 
resolved study, while the solid lines at the bottom indicate the occurrence time 
of the bursts.} 
\label{LCcha}
\end{figure}

The \nustar\ (Harrison et al. 2013) observation (obsid 90201017002) 
started on 2016-04-21 at 14:41:08 UT. The data were reduced with the 
standard {\it nupipeline} scripts v. 0.4.5 and the high level products 
produced with the {\it nuproducts} tool. 
The \chandra\ spectra (obsid 17858 on 2016-04-21 13:44:43 UT) 
and response matrices have been 
produced with the {\sc chandra\_repro} task, combining the positive 
and negative first orders. 
The light curve was extracted with the {\sc dmextract} task. 
Bursts were singled out by visually inspecting the \chandra\ and \nustar\ 
light curves and selecting intervals of enhanced emission (typically 
lasting $\approx100-200$~s; Fig. \ref{LCcha}). 

Figure \ref{spec} shows the mean \chandra\ and \nustar\ spectra. 
As hoped, we caught the source during the soft state. The simultaneous 
spectra show, in addition to the typical soft state continuum, an array 
of more than 60 absorption lines. The absorption lines are 
due to highly ionised plasma (Ponti et al. in prep). 
The strongest lines are labeled in 
Fig. \ref{spec} and they correspond to the Ly$\alpha$ transitions 
(as well as the \Fevc\ line) of the most abundant elements. 
The broad band continuum can be fit by the sum of a disk black-body 
($kT_{DBB}\approx1.1-1.5$~keV), black-body ($kT_{BB}\approx2.5-3.0$~keV) 
and a Comptonisation component, all absorbed by neutral material 
($N_H$$\approx$$1.5-2.1$$\times$$10^{21}$ cm$^{-2}$; for more details 
Ponti et al. in prep). 

Figure \ref{LCcha} shows the first order \chandra\ light curve. 
Two eclipses are detected, preceded by intense dipping activity. 
By fitting the eclipse transitions, we determined the eclipse center 
with an accuracy of seconds ($P_{\rm orb}$$=$$25.618$~ks, 
consistent with previous results; Jain et al. 2017; Iaria et al. 2018). 
We first divided the dataset into intervals of $3659.7$~s, so that 
7 intervals cover an entire orbital period. 
The start of the first interval is chosen so that 
it begins just after the end of the eclipse (see grey dotted 
lines in Fig. \ref{LCcha}; we define as phase 0 the eclipse center). 
We then removed the periods affected 
by bursts and eclipses. This resulted in a shorter cleaned 
exposure for the $7^{th}$ interval, because of the presence of the eclipses. 
We then extracted the HEG and MEG first order spectra 
corresponding to the first intervals after the eclipses 
(accumulating the spectra over both orbital periods), and 
so forth.

\begin{figure}
\begin{center}
\vspace{-0.3cm}
\includegraphics[height=0.29\textwidth,angle=0]{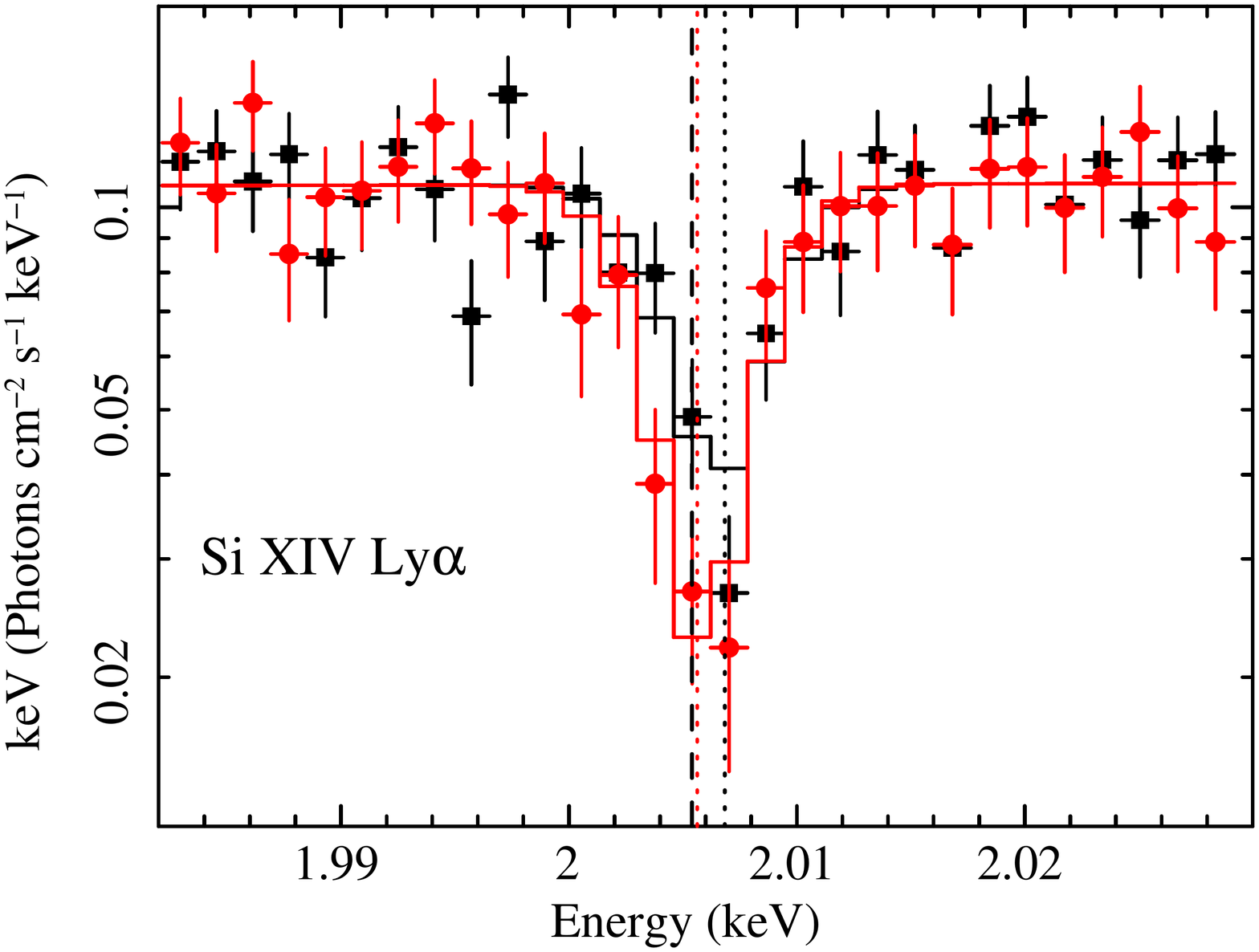}

\vspace{-0.6cm}
\includegraphics[height=0.29\textwidth,angle=0]{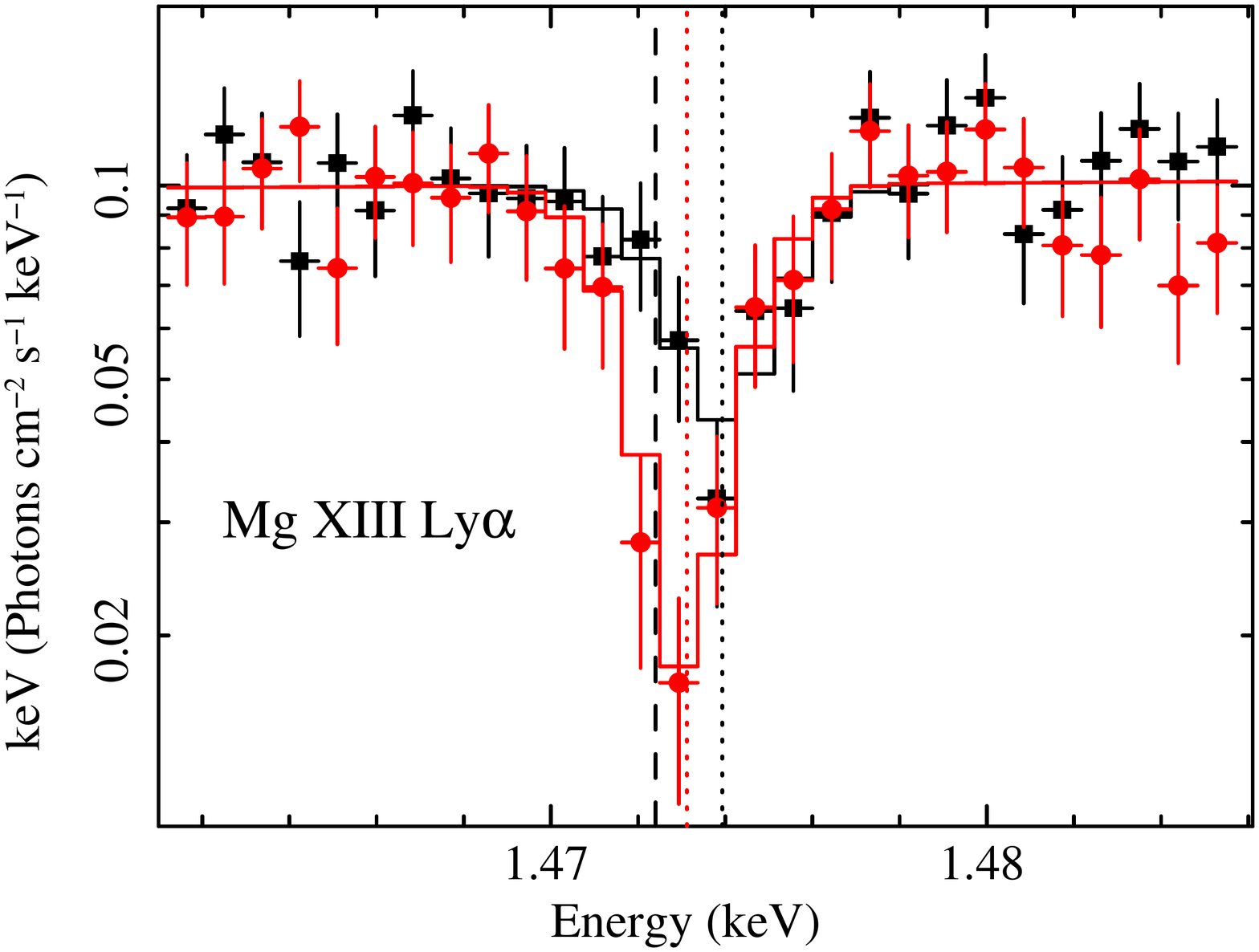}
\vspace{-0.3cm}
\caption{{\it (Top panel)} Black squares, red circles and solid lines 
show the MEG spectra and best fit models at the energy of the 
Si~{\sc xiv} line, accumulated during phase 0.02-0.16 and 
0.59-0.73, respectively. The dotted lines indicate the best fit Gaussian 
line energies, while the dashed line show the expected line transition. 
{\it (Bottom panel)} Spectra and best fit models around the 
Mg~{\sc xii} transition, accumulated during phase 0.16-0.31 and 
0.31-0.45, respectively. The vertical lines carry the same meaning 
as the top panel. }
\end{center}
\label{CPhase}
\end{figure}

\vspace{-0.7cm}
\section{Phase dependent absorption variations} 
\label{Pdep}

We started by simultaneously fitting the HEG and MEG spectra 
within a narrow energy band ($\Delta\lambda/\lambda\approx0.06$) 
centred on the absorption feature under consideration. 
We applied this process for the four strongest soft absorption 
lines: Mg~{\sc xii} Ly$\alpha$ ($\lambda_0=8.4210$~\AA); 
Si~{\sc xiv} Ly$\alpha$ ($6.1822$); S~{\sc xvi} Ly$\alpha$ 
($4.7292$) and Ar~{\sc xviii} Ly$\alpha$ ($3.7329$~\AA)\footnote{The 
reported wavelengths correspond to the average of the wavelengths 
of the respective doublets (Mg~{\sc xii}: 8.4192, 8.4246; Si~{\sc xiv}: 
6.1804, 6.1858; S~{\sc xvi}: 4.7274, 4.7328; Ar~{\sc xviii}: 3.7311, 
3.7365~\AA), weighted over the oscillator strength. We repeated 
the analysis and we considered two Gaussian lines for the doublets, 
obtaining the same results.}. 
We fitted the continuum with the best fit model of the mean 
\chandra+\nustar\ spectra, leaving only the normalisation 
of the disk black-body component and the column density 
of the neutral absorber free to vary as a function of 
phase\footnote{We note that, due to the small energy band 
considered, a similar result is obtained by fitting the phase 
resolved spectra with a simple power law, with free normalisation
only.}. 

Figure \ref{LCcha} shows that the spectra during interval 6 and 7 
are affected by intense dipping activity. 
Dipping is typically generated by an increased 
column density of low ionisation absorption, causing a drop 
of the soft X-ray continuum flux (Sidoli et al. 2001; 
Boirin et al. 2004; Diaz-Trigo et al. 2006). 
This effect is captured in our narrow band fits by either 
an increase in the column density of the neutral absorber or 
a drop of the normalisation of the continuum (e.g., during interval 6). 
Because of the presence of the dips and the shorter cleaned 
exposure, we do not consider interval 7 here. 

We independently fitted each strong absorption line 
as a function of phase (e.g., from interval 1 to 6; 
phases 0.02 to 0.88). We performed this by adding 
to the continuum a Gaussian absorption profile, 
with energy and intensity of the line free to vary as a function 
of phase (after verifying that the line width was consistent 
with being constant). 
\begin{figure}
\vspace{-0.4cm} 
\hspace{-0.5cm}
\includegraphics[height=0.43\textwidth,angle=0]{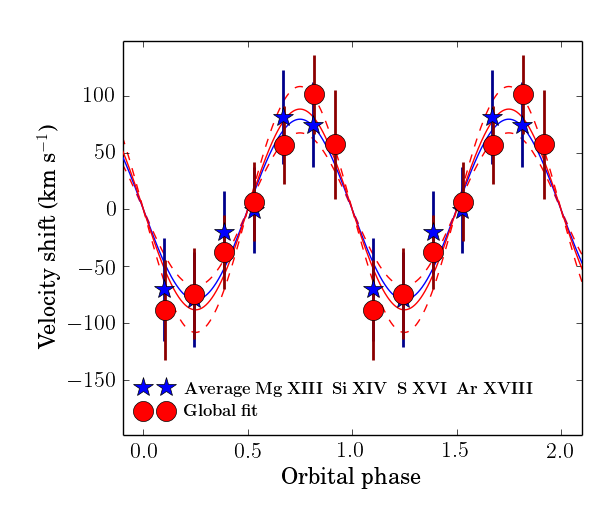}

\vspace{-0.5cm}
\caption{The blue stars and red dots show the average velocity 
shift for the Mg~XII~Ly$\alpha$, Si~XIV~Ly$\alpha$, S~XVI~Ly$\alpha$ 
and Ar~XVIII~Ly$\alpha$ lines and for all the lines, respectively 
(1-$\sigma$ statistical errors). 
The lines appear red-shifted at phase 
$\approx$$0.25$ and blue-shifted at $\approx$$0.75$, as expected by the 
movement of the NS around the centre of mass of the binary. 
The blue and red lines show the best fit with a sinusoid 
for the average four lines ($K_1$$=$$80\pm22$ km s$^{-1}$)
and all the lines ($K_1$$=$$89\pm19$ km s$^{-1}$), respectively. 
The red dashed lines indicate the 1-sigma uncertainties on 
the best fit model. An offset smaller than the instrument calibration 
($\gamma$$=$$-49$~km~s$^{-1}$) has been subtracted off.}
\label{Shifts}
\end{figure}
For all lines, we observed that the line centroid energies showed 
a larger blue shift in the first part of the orbit, compared with later on
(e.g., see Fig.\ 3).
To measure the amplitude and significance of this effect, 
we recorded for each of the strongest soft X-ray lines the velocity 
shift and its error, as a function of phase. The blue stars 
in Fig. \ref{Shifts} show, as a function of phase, the weighted 
average of the observed shifts for the four strongest soft lines 
(Mg~{\sc xii}; Si~{\sc xiv}; S~{\sc xvi} and Ar~{\sc xviii}).
A fit with a constant velocity provides a best fit value of $\gamma=-48$~km~s$^{-1}$
and a $\chi^2=14.2$ for 5 dof. We then added to the model 
a sinusoid component with free amplitude and zero-phase ($\varphi_\mathrm{0}$; 
the period was assumed to be unity). 
The best fit yielded $\varphi_\mathrm{0}=0.54\pm0.05$, consistent 
with an orbital modulation induced by the motion of the primary 
($\varphi_\mathrm{0}$=0.5). Therefore, we re-fitted 
the data leaving only the semi-amplitude (i.e. $K_\mathrm{1}$) as free 
parameter (i.e. fixing $\varphi_\mathrm{0}$=0.5). This significantly 
improved the fit ($\Delta\chi^2=12.7$ for 
the addition of one parameter, corresponding to a $\approx3\sigma$ 
improvement). The observed best fit values are: 
$\gamma=-49\pm16$~km~s$^{-1}$ and 
$K_\mathrm{1}=80\pm22$~km~s$^{-1}$ (Fig. 
\ref{Shifts})\footnote{To estimate the uncertainties on the best fit 
parameters, we simulated $10^5$ radial velocity curves 
where each value of the velocity is due to a randomisation 
of the observed velocity assuming a normal distribution with 
a width as large as its observed uncertainty. 
Then, for each randomisation, we computed the best fit 
sinusoidal function. Finally, from the envelope of $10^5$ 
best fit functions, we determined the uncertainties on the best fit 
parameters from the lower 15.9 and upper 84.1 
percentiles on the envelope of the functions. }.
The observed $\gamma$ might, in theory, trace a bulk outflow 
velocity of the ionised plasma, such as a wind, or be related with 
the systemic velocity of \mxb.  
However, the absolute wavelength accuracy of the HEG 
is $\pm0.006$~\AA, corresponding to $\approx140$~km~s$^{-1}$ at 1 keV
($\sim3$ times larger than $\gamma$), therefore, we do not discuss 
$\gamma$ any further. 
\begin{figure*}
\hspace{-0.6cm}
\includegraphics[height=0.295\textwidth,angle=0]{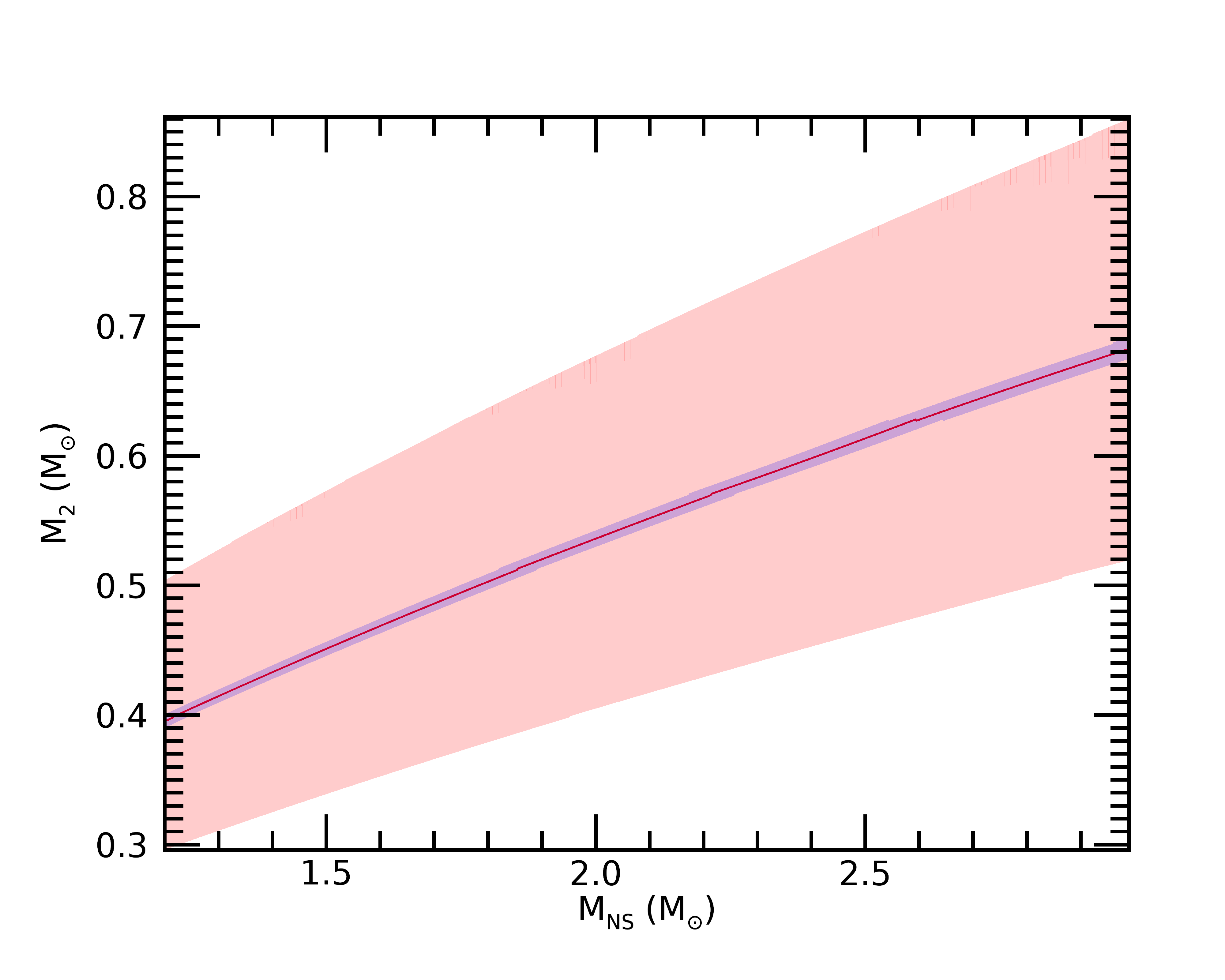}
\hspace{-0.6cm}
\includegraphics[height=0.295\textwidth,angle=0]{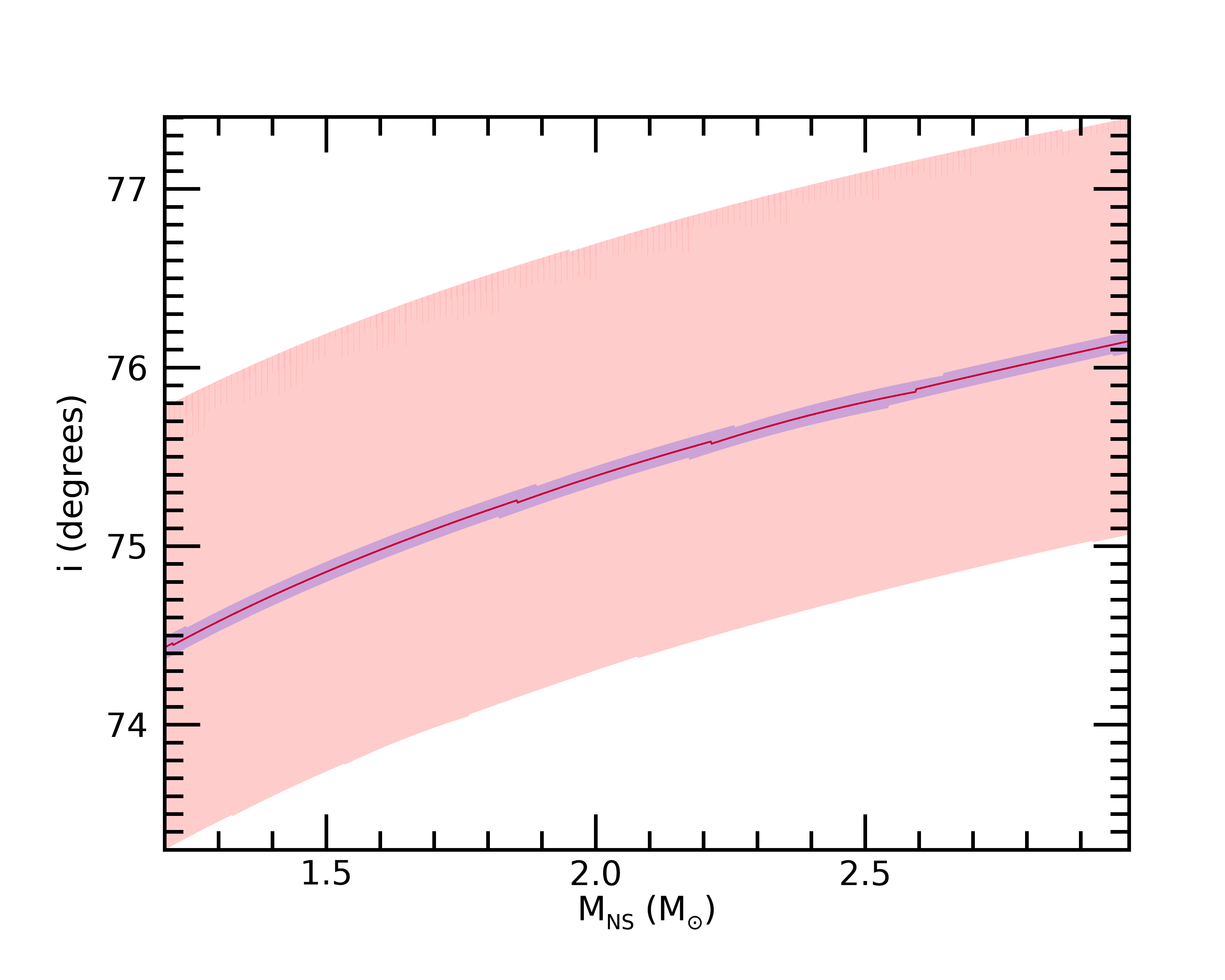}
\hspace{-0.6cm}
\includegraphics[height=0.28\textwidth,angle=0]{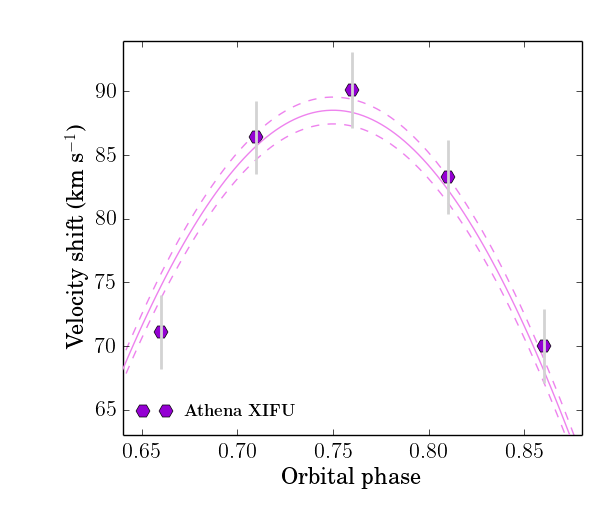}
\vspace{-0.45cm}
\caption{{\it (Left and center panels)} 
Constraints on the orbital inclination ($i$), NS ($M_{NS}$) and 
companion star masses ($M_2$) derived from the mass function (eq. 1) 
and the eclipse duration (eq. 3), under the assumption that the companion 
star fills its Roche lobe (eq. 4). The pink and violet areas show the region 
allowed by the current and future (\athena) uncertainties.   
{\it (Right panel)} Determination of the semi-amplitude of the radial 
velocity curve through a 100~ks \athena\ XIFU observation. 
A zoom on a small part of the radial velocity curve is shown, in order 
to exhibit the small uncertainties. 
$K_\mathrm{1}$ will be precisely constrained to 
($K_\mathrm{1}=89.0\pm0.9$~km~s$^{-1}$).}
\label{Mass}
\label{XIFUShifts}
\vspace{-0.3cm}
\end{figure*}

To improve the determination of the velocity shift, we built 
a self-consistent photo-ionisation model ({\sc IA$_{\rm soft}$}). 
The model table was computed with {\sc cloudy} 17.00 (Ferland et al. 
2013), providing as input the observed soft state spectral 
energy distribution, constant electron density 
$n_e=10^{14}$~cm$^{-3}$, turbulent velocity 
$v_{turb}=500$~km~s$^{-1}$ and Solar abundances (for more 
details see Ponti et al. in prep). 
We then fitted the phase resolved spectra, over the entire 
usable energy range (0.8-6~keV for MEG and 1.2 -7.3~keV for HEG).
The advantage of this global fit, compared with the independent 
fit of each single line, is that the energy separation between the different 
absorption lines is set by the atomic physics in the model. 
We fitted the spectra with a disk black-body ({\sc diskbb}) plus black-body 
({\sc bbody}) model absorbed by neutral ({\sc tbabs}) plus 
ionised material (in {\sc Xspec} jargon: {\sc tbabs*IA$_{\rm soft}$*(diskbb+bbody)}; 
Arnaud 1996), with all parameters free to vary\footnote{Thanks to the 
better determination of the continuum allowed by the broad band fit, 
we included also interval 7 to this analysis.}. 
The red points in Fig. \ref{Shifts} show the best fit velocity shift, 
as a function of phase. We note that the consideration of the full array 
of absorption lines led to results fully consistent with those 
obtained by analysing only the four strongest lines and 
to slightly reduce the error bars on the radial velocity measurement. 

The fit with a constant could be rejected ($\chi^2=21.3$ for 6 dof). 
The fit significantly improved by adding a sinusoidal 
component\footnote{The best fit phase resulted to be $0.01\pm0.04$, 
therefore we re-fitted the data, fixing the phase to be equal to 0 in the fit.} 
($\Delta\chi^2=19.3$ for the addition of one parameter, corresponding 
to $\sim3.5\sigma$ improvement), producing an acceptable fit ($\chi^2=2.0$
for 5 dof). The best fit yielded $K_\mathrm{1}=89\pm19$~km~s$^{-1}$ 
(see red solid and dashed lines in Fig. \ref{Shifts}).
The HEG relative wavelength accuracy is $\pm0.001$~\AA, 
corresponding to $\approx25$~km~s$^{-1}$ at 1 keV, comparable to the statistical 
uncertainties on the velocity shift at each orbital phase. 
Therefore, our measurement of $K_\mathrm{1}$ is solid. 

\vspace{-0.6cm}
\section{Discussion}

Direct measurements of $K_\mathrm{1}$ have been possible 
for the case of X-ray binary pulsars, where the delay in the arrival 
time of the pulses accurately trace the orbit of the primary and 
therefore $K_\mathrm{1}$. 
As an alternative, we used the absorption lines from ionised material likely formed 
in the atmosphere of the inner accretion disc. 
Since they arise from a confined region surrounding the compact object 
they are expected to trace its motion around the centre of mass 
of the binary (Zhang et al. 2012). The measured $\varphi_\mathrm{0}$ 
of the sinusoidal radial velocity curve corroborates this picture. 
Likewise, one can compare the derived $K_\mathrm{1}=89\pm19$ km s$^{-1}$ 
with that determined in other NS LMXB with similar orbital periods and inclinations. 
We observed that our $K_\mathrm{1}$ measurement is very similar to those 
derived in these other systems. Indeed, both the eclipsing LMXB X-ray pulsar X1822-371 
(orbital period $P_{\rm orb}=5.6$ hr; Hellier \& Mason 1989) and EXO~0748-676
($P_{\rm orb}=3.8$ hr; Parmar et al. 1986) have 
$K_\mathrm{1}=94.5\pm0.5$ km s$^{-1}$ and 
$K_\mathrm{1}=100\pm20$ km s$^{-1}$, respectively (Jonker \& Van der Klis 2001;
Mu\~noz-Darias et al. 2009; although the latter value is derived from optical data via 
diagnostic diagrams therefore less accurate; Shafter et al. 1986). 
On the contrary, black holes and low-inclination NS systems have typically 
significantly smaller $K_\mathrm{1}$ values (e.g. Casares \& Jonker 2014). 
Since both the phase offset and velocity amplitude derived from 
the absorption lines energy shifts are consistent with tracing the motion 
of the NS, they can be used to constrain some of the fundamental 
parameters of \mxb.

\vspace{-0.5cm}
\subsection{Constraints on the NS mass and the orbital inclination}

The measured radial velocity curve of the NS allowed us to derive 
the mass function of the system: 
\begin{equation}
\frac{K_1^3 P_{orb}}{2\pi G} = \frac{M_2^3 sin^3 i}{(M_{NS}+M_2)^2},
\end{equation}
where $M_{NS}$ and $M_2$ are the NS and the companion star masses, 
$G$ the Gravitational constant and $i$ the orbital inclination. 
This equation sets the first constraint on the unknowns: $M_{NS}$, $M_2$ and $i$. 

The knowledge that the eclipse duration lasts $\Delta T_{ecl}=899.1\pm0.6$ 
(Iaria et al. 2017) adds an additional constraint. 
Indeed, following Iaria et al. (2017), we have computed the size of 
the occulted region $x$ (see fig. 5 in Iaria et al. 2017) as: 
\begin{equation}
x=\frac{ \pi a \Delta T_{ecl} } {P_{orb}},
\end{equation}
where $a$ is the orbital separation ($a^3=\frac{G(M_{NS}+M_2)}{4\pi}$). 
This allowed us to constrain $i$ as a function of $M_{NS}$ and $M_2$: 
\begin{equation}
tan^2(i)= \frac{R_2^2-x^2}{a^2-(R^2_2-x^2)},
\end{equation}
where $R_2$ is the companion star radius, by assuming that 
the secondary star is filling its Roche lobe ($R_2=R_L$; Paczy\'nsky 1971):
\begin{equation}
R_2=R_{L2}=0.462 a (\frac{M_2}{M_{NS}+M_2})^{1/3}.
\end{equation}

By applying these two constraints to the three unknowns, we estimated that, 
for any reasonable NS mass ($1.2\leq M_{NS}\leq 3$~M$_\odot$), the 
companion star mass should lie within the range 
$0.3\leq M_2\leq 0.8$~M$_\odot$ and the orbital inclination 
$73 \leq i \leq 77^\circ$ (Fig. \ref{Mass}).
The measured ranges of the most likely companion star masses 
and orbital inclinations are consistent with previous methods (Wachter et al. 2000). 
This demonstrates that it is possible to constrain the fundamental 
parameters of \mxb\ with current X-ray data and that this method can be 
applied to other systems. 

\vspace{-0.5cm}
\subsection{Limitations and relevance of the method}

The technique employed here presents also limitations. 
For example, it requires the presence of a further constraint 
(e.g., via optical-nIR observations; Casares \& Jonker 2014), 
in order to eliminate the degeneracy between the orbital inclination and
the masses of the two stars. 
Additionally, it assumes that the bulk motion of either the accretion disc 
wind or of the ionised disc atmosphere is azimuthally symmetric and constant 
during the orbit. Despite this assumption appears reasonable at first approximation, 
second order effects might be present. For example, the wind outflow 
velocity might vary, implying that the radial velocity curve should 
be averaged over several orbital periods. Besides, the wind/atmosphere 
is likely structured into a multi-phase plasma, characterised by different 
physical parameters, possibly complicating the analysis.  
Additionally, the wind/atmosphere kinematic might be perturbed 
by the material transferred from the companion star or by the presence 
of disc structures (e.g., eccentric discs; strong warps). 

Currently the main limitation is due to the large statistical uncertainties 
on the determination of the radial velocity curve. 
However, the next generation of X-ray spectrographs/calorimeters 
will mend this state of affairs. Indeed, compared with HETG, 
\xarm\ and \athena\ will improve the resolving power and 
the figure of merit for weak line detections at 6~keV, by a factor 
of $\approx6$, $\approx12$ and $\approx8$, $\approx40$, respectively, 
while \arcus\ at 1.5~keV will improve it by $\approx3$ and $\approx7$,
respectively (Nandra et al. 2013; Kaastra et al. 2016; Gandhi 2018). 
We also note that this method can be applied to any LMXB 
displaying ionised absorption lines produced by the disc atmosphere/wind.
This category comprises the majority of the high inclination LMXB 
(about half of the sample) during their softer states (Diaz-Trigo et al. 2006; 
Ponti et al. 2012). 

\vspace{-0.6cm}
\subsection{The power of \athena-XIFU to constrain masses}

To measure the power of future X-ray spectroscopy to constrain the mass 
function of \mxb, we simulated an \athena-XIFU 100~ks 
observation (assuming the "as proposed" version of the 
response matrix with reduced effective area, to preserve excellent 
spectroscopy at high throughput; Barret et al. 2016). 
We assumed the same observed flux 
($F_{0.5-10~keV}=9.5\times10^{-10}$~erg~cm$^{-2}$~s$^{-1}$), 
spectral continuum and ionised plasma parameters as measured 
during the \chandra\ observation (Ponti et al. in prep). We further 
assumed that the ionised plasma is affected by the radial velocity 
curve of the primary, as observed. Figure \ref{XIFUShifts} demonstrates 
that \athena\ observations will allow us to determine the amplitude 
of the radial velocity curve with an uncertainty of $\approx$$1$~\%. 
The blue curves in Fig. \ref{Mass} show the constraints that such 
an observation will allow us to deliver, translating into an uncertainty 
of $\approx$$5$~\% on the mass of the primary, would the radial velocity 
of the companion star be known (e.g., via optical spectroscopy). 
Therefore, future X-ray missions will be able to deliver measurements 
of the radial velocity of LMXB to an accuracy of a few per cent.

\vspace{-0.6cm}
\section*{Acknowledgments}

{The authors wish to thank Fiona Harrison, Karl Forster and the \chandra\ team 
for approving the \nustar\ DDT and quickly scheduling the simultaneous observations.
The authors also wish to thank Eugene Churazov, Joachim Tr\"umper and Florian 
Hofmann for discussion. 
GP acknowledges financial support from the Bundesministerium f\"{u}r Wirtschaft 
und Technologie/Deutsches Zentrum f\"{u}r Luft- und Raumfahrt 
(BMWI/DLR, FKZ 50 OR 1604, FKZ 50 OR 1715) and the Max Planck Society. 
SB acknowledges financial support from the Italian Space Agency under 
grants ASI-INAF I/037/12/0 and 2017-14-H.O.
TMD acknowledges support by the Spanish MINECO via a Ram\'on y Cajal 
Fellowships (RYC-2015-18148), the grant AYA2017-83216-P and 
the EU COST Action CA16214 (STSM reference number: 40355).}

\vspace{-0.6cm}
{}

\end{document}